\documentclass[preprint]{aastex}

\begin{document}

\title{The contribution of the kinematic Sunyaev--Zel'dovich Effect 
from the Warm Hot Intergalactic Medium  
to the Five-Year {\it Wilkinson Microwave Anisotropy Probe} Data} 

\author{R. G\'enova-Santos\altaffilmark{1,2}, F. Atrio-Barandela\altaffilmark{3}, 
J.P. M\"ucket\altaffilmark{4} \& J.S. Klar\altaffilmark{4}}
\altaffiltext{1}{Instituto de Astrof\'{\i}sica de Canarias, 
V\'{\i}a L\'actea, s/n. 38200 La Laguna, Tenerife, Spain; email:rgs@ll.iac.es}
\altaffiltext{2}{Astrophysics Group, Cavendish Laboratory, 
University of Cambridge CB3 OHE, UK}
\altaffiltext{3}{F\'{\i}sica Te\'orica, Universidad de Salamanca, 
37008 Salamanca, Spain; email: atrio@usal.es}
\altaffiltext{4}{Astrophysikalisches Institut Potsdam. D-14482 Potsdam, Germany; 
email: jpmuecket@aip.de; email: jklar@aip.de}

\begin{abstract}
We study the contribution of the kinematic Sunyaev--Zel'dovich (kSZ) effect, generated by
the warm-hot intergalactic medium (WHIM), to the cosmic microwave background (CMB)
temperature anisotropies in the Five-Year {\it Wilkinson Microwave 
Anisotropy Probe} (WMAP) data. We explore the concordance $\Lambda$CDM cosmological model, 
with and without this  kSZ contribution, using a Markov chain Monte Carlo algorithm. 
Our model requires a single extra parameter to describe this new component.
Our results show that the inclusion of the kSZ signal improves the fit to the data 
without significantly altering the best-fit cosmological parameters except 
$\Omega_{\rm b}h^2$. The improvement is localized at the $\ell\gtrsim 500$ multipoles. 
For the best-fit model, this extra component peaks at $\ell\sim 450$ with an amplitude  
of $129~\mu$K$^2$, and represents 3.1\% of the total power measured by the
{\it Wilkinson Microwave Anisotropy Probe}. 
Nevertheless, at the 2$\sigma$ level a null kSZ contribution is still compatible with 
the data. Part of the detected signal could arise from unmasked point sources 
and/or Poissonianly distributed foreground residuals.  A statistically more significant 
detection requires the wider frequency coverage and angular resolution of the 
forthcoming {\it Planck} mission.
\end{abstract}

\section{Introduction}
Baryons represent a small fraction of the total mass--energy budget of 
the Universe and do not play a predominant role in its evolution. 
They are the only matter component that has been identified directly.
The baryon fraction of the Universe has been determined at different redshifts
through a variety of methods: $\Omega_{\rm b}h^2=0.020\pm 0.002$ \citep{burles_01} 
from Big Bang nucleosynthesis (BBN), $\Omega_{\rm b}h^2>0.021$ \citep{rauch_97} from 
the Ly$\alpha$ forest and $\Omega_{\rm b}h^2=0.02273\pm 0.00062$ \citep{dunkley_09} 
from the cosmic microwave background (CMB) primary anisotropies. 
The baryon fraction measured from the  well-observed components at $z=0$ is 
$\Omega_{\rm b}h^2 = 0.010\pm 0.003$ \citep{fukugita_98}, indicating that half of the baryons in 
the local Universe are still undetected.  

Cosmological simulations of large-scale 
structure formation \citep{petitjean_95,zhang_95, hernquist_96, 
katz_96,theuns_98,dave_99} have shown that the
intergalactic gas has evolved from the initial density perturbations
into a complex network of mildly nonlinear filaments in the
redshift interval $0 < z < 6$. With cosmic evolution, a significant
fraction of the gas collapses into bound objects; baryons in  
the intergalactic medium (IGM) are in filaments containing Ly$\alpha$ systems 
with low HI column densities and, at low redshifts, shock-confined gas with 
temperatures $T_{\rm e}\sim 0.01$--1~keV and overdensities $\delta_{\rm b}\sim 10-50$ 
\citep{dave_01,cen_06}.
An important fraction of the missing baryons could be located in 
this web of shock-heated filaments, called ``warm/hot intergalactic medium'' 
(WHIM).  Observational efforts to detect this ``missing baryon'' component range from 
looking at its emission in the soft X-ray bands \citep{zappacosta_05}, 
ultraviolet absorption lines in the spectra of more distant sources \citep{nicastro_05}
or Sunyaev--Zel'dovich (SZ) contributions in the direction of superclusters of 
galaxies (G\'enova-Santos et al. 2008; see Prochaska \& Tumlinson 2008 for a review). 
Indirect searches of the WHIM using the SZ effect have been inconclusive. The SZ imprint 
due to galaxy clusters in {\it WMAP} data is well measured \citep{atrio_08b}, but when this 
component is removed no signal associated to the WHIM remains \citep{hernandez_04}.

\citet{atrio_06} developed a formalism to account for the contribution 
of the IGM/WHIM to the CMB anisotropies via the thermal SZ (tSZ, Sunyaev 
\& Zeldovich 1972) effect. The main assumption was that missing baryons were 
distributed as a diffuse gas phase outside bound objects. Its filamentary structure 
was assumed to be described by a log--normal distribution function, which accurately models 
a mildly non-linear density field when the velocity field remains in the
linear regime \citep{coles_91}. The predicted power spectrum of the tSZ effect peaks at
$\ell\sim$~2000--4000, with an amplitude similar to the tSZ from galaxy clusters. 
More recently, in \citet{atrio_08}  we studied the contribution of the 
kinematic SZ (kSZ, Sunyaev \& Zeldovich 1980) effect. We found that the kSZ power spectrum 
had a maximum at $\ell\sim$~400-600. In this article we search for a possible kSZ 
contribution in the 5 year {\it Wilkinson Microwave Anisotropy Probe} ({\it WMAP}) data.
We use a Markov chain Monte Carlo (MCMC) method to sample the parameter space 
of the concordance $\Lambda$CDM model with and without a kSZ component to determine
if there is a statistically significant contribution. Briefly,
in Sections~2 and 3 we describe the model and the numerical implementation 
of the MCMC, and in Section~4 we present our results and summarize our main conclusions. 

\section{The thermal and kinematic SZ effect from the IGM/WHIM.}

The tSZ effect is the weighted average of the electron pressure along
the line of sight; the kSZ is proportional to the column density of
the free electrons along the line of sight $\hat{n}$, weighted by the 
radial component of their peculiar velocities:
\begin{equation}
\left(\frac{\Delta T}{T_0}\right)_{\rm tSZ}(\hat{n})=G(\nu)\frac{k_{\rm B}\sigma_{\rm T}}{m_{\rm e} c^2}
\int dl\; n_{\rm e} T_{\rm e}\; ,\qquad \qquad
\left(\frac{\Delta T}{T_0}\right)_{\rm kSZ}(\hat{n})=\frac{\sigma_{\rm T}}{c}
\int dl\; n_{\rm e} (\vec{v}_{\rm e}\cdot\hat{n})\; .
\end{equation}
In these expressions, $T_{\rm e}$, $n_{\rm e}$, $v_{\rm e}$ are the electron temperature,
density and peculiar velocity, respectively, 
$k_{\rm B}$ is Boltzmann constant, $\sigma_{\rm T}$ Thompson cross section, 
$m_{\rm e}c^2$ the electron annihilation energy, 
$c$ is the speed of light and $G(\nu)$ is the frequency dependence of
the tSZ effect. In the Rayleigh--Jeans regime, $G(\nu)\approx -2$ with less than 
20\% variation at {\it WMAP} frequencies. For an isothermal cluster
the tSZ to kSZ ratio is: $(\Delta T_{\rm tSZ}/\Delta T_{\rm kSZ})\simeq
20 G(\nu)(T_{\rm e}/10\,{\rm keV})(300\,{\rm km\,s^{-1}}/v_{\rm e})$. The temperature of 
the IGM is much lower than in galaxy clusters even in the shock-heated WHIM and the kSZ 
contribution could become comparable to that of the tSZ effect.

In \citet{atrio_06} and \citet{atrio_08}  we computed the 
tSZ and kSZ temperature anisotropies generated by the IGM assuming 
that baryons are distributed as in a log--normal random field.
The log--normal distribution was introduced by Coles \& Jones (1991)
as a model for the non-linear distribution of matter in the Universe.
The number density of electrons $n_{\rm e}$ can be obtained from
the baryon distribution assuming ionization equilibrium between recombination 
and photoionization and collisional ionization. In the
conditions valid for the photoionized IGM the gas is almost completely ionized.
The correlation function of the tSZ temperature anisotropy generated by two 
filaments located at two different redshifts along 
two lines of sight with an angular separation $\alpha$ is 
dominated by the spatial variations of the electron
pressure at nearby locations \citep{atrio_06}. Within
the small angle approximation:
\begin{equation}
C(\alpha)=\left[\frac{k_{\rm B}\sigma_{\rm T}}{m_{\rm e}c^2}G(\nu)\right]^2
\int_0^{z_f}dz{\left(\frac{dl}{dz}\right)^2 n^2_e(z)T^2_0(z)
e^{\gamma(\gamma-1)\Delta^2(z)}[e^{\gamma^2Q(\alpha,z)}-1]}~~.
\label{eq:tsz}
\end{equation}
In this expression, $\Delta^2(z)$ is the variance of the baryon density field, 
which is related to the dark matter power spectrum $P(k)$ by:
\begin{equation}
\Delta^2(z)=D^2(z)\int{\frac{d^3k}{(2\pi)^3}\frac{P(k)}{[1+x_b^2(z)k^2]^2}}
\label{eq:deltaB}
\end{equation}
where $D(z)$ is the growth factor of matter density perturbations and 
$x_b$ is the comoving Jeans length. Also,
\begin{equation}
Q(\alpha,z)=\frac{D^2(z)}{2\pi^2}
\int^\infty_0\frac{P(k)k^2 dk}{[1+x_b^2(z)k^2]^2}j_0(k\alpha)~~,
\label{Q}
\end{equation}
where $j_0$ is the zeroth order spherical Bessel function.
The integration in eq.~(\ref{eq:tsz}) extends to the highest redshift $z_f$.

Similar arguments can be used to compute the correlation function
of the kSZ effect due to two filaments located at distances 
$l$ and $l'$ along two lines of sight separated by an angle $\alpha$
(see \citet{atrio_08}  for further details): 
\begin{equation}
C(\alpha)=f_{\rm b}^2\frac{\sigma_{\rm T}^2}{c^2}\int_o^{l(z_f)}\int_o^{l'(z_f)} dl dl' 
V_{\rm B}(R(z))D_v(z)V_{\rm B}(R(z'))D_v(z') \langle n_{\rm e}(l\hat{n},z)n_{\rm e}(l'\hat{n}',z')\rangle ~~,
\label{eq:ksz}
\end{equation}
where $V_{\rm B}(R(z))$ denotes the mean bulk velocity of a sphere with electron 
density $n_{\rm e}$ and radius $R(z)$, the comoving distance from the observer to a 
filament at redshift $z$, $D_v(z)$ is the velocity linear growth factor
and $f_{\rm b}$ is the fraction of the baryons in WHIM filaments. 

The correlation functions given in eqs~(\ref{eq:tsz}) 
and (\ref{eq:ksz}) differ in two significant aspects:
(a) While the tSZ effect is cumulative and several filaments along
the line of sight add linearly to the total effect, eq.~(\ref{eq:ksz}) requires
all filaments to be moving with the same velocity. This is the physical
reason why only the bulk flow velocity contributes to the effect. If 
there are several filaments, in a given direction, with different velocities the net
effect will not be $\propto (n_{\rm e}V_{\rm B})$ but $\propto (n_{\rm e}V_{\rm B})^{1/2}$.
(b) The IGM is usually described by a polytropic 
equation of state $T\propto n^{\gamma-1}$, i.e.,
high dense regions contribute more
to the tSZ effect since their temperature is higher. In eq.~(\ref{eq:tsz}) 
we then integrate all scales out to the largest overdensity $\delta_{\rm max}$ that 
is well described by the log--normal model. Otherwise we would be including
small contributions from gas in bound objects and not described by the log--normal model.
This restriction is not necessary for the kSZ effect.  In the log--normal 
approximation, velocities are in the linear regime and they are not
correlated with matter overdensities. Then, instead of introducing an
arbitrary cut-off at overdensity $\delta_{\rm max}$, 
we extend our integration to all overdensities, i.e., to
all baryons, and introduce a factor $f_{\rm b}$ to take into account 
the actual fraction of baryons in the WHIM; we could then, in principle,
constrain this fraction directly from the data.

Eqs~(\ref{eq:tsz}) and (\ref{eq:ksz}) can be inverted to give respectively the 
radiation power spectra $C_\ell^{\rm tSZ}$ and 
$C_\ell^{\rm kSZ}$ at each multipole $\ell$. These calculations are computationally 
expensive since, to get accurate results, we use line-of-sight separations of 30$''$ 
and redshift intervals of $\Delta z=0.005$ up to $z=0.1$ and $\Delta z=0.1$ 
up to $z_f=1$, where the contribution becomes negligible. The final tSZ
power spectrum depends on cosmological and physical parameters: 
the cut-off scale $\delta_{\rm max}$, the amplitude 
of the matter density fluctuations on spheres of $8h^{-1}$~Mpc,
$\sigma_8$, the mean gas temperature, $T_0$,  the 
gas polytropic index, $\gamma$, and the mean gas temperature at reionization, 
$T_m$. The kSZ contribution depends on the Jeans length $x_b$,
$\sigma_8$ and $f_{\rm b}$. Since the product $\gamma T_m$ fixes $x_b$ then
the kSZ contribution requires less parameters than the tSZ.
  
In Figure~\ref{fig1}a we compare CMB power spectrum of the concordance 
$\Lambda$CDM model with the kSZ and tSZ WHIM contributions. The parameters of 
the concordance model are those of the best-fit 5 year {\it WMAP} data.
The SZ contributions are calculated using: $\sigma_8=0.77$, $\gamma=1.3$, 
$\gamma T_m\approx 1.3\times 10^4$~K  and $f_{\rm b}=0.5$.
First, the relative amplitude of the thermal and kinematic contributions
depends on model parameters; the kSZ effect could be larger than the
tSZ effect since the average temperature of the IGM is rather low. 
Second, since the IGM is not isothermal, the tSZ contribution is dominated by
the mildly overdense regions, which subtend smaller angles than the filaments themselves,
so its power is shifted to higher $\ell$. 
In Figures~\ref{fig1}b,c we show the contribution of different
redshift intervals to the total power for each of the WHIM contributions.
In both cases, the main contribution comes from $z=0$ to $0.4$. 
Contributions from higher redshifts decrease rapidly.

The purpose of this article is to search for a IGM/WHIM contribution in the 5 year 
{\it WMAP} data, which are sensitive to $\ell\lesssim 1000$ multipoles. On these angular
scales, the kSZ contribution is largest while the tSZ is larger at $\ell\sim 3000$.
\citet{friedman_09} see no evidence of a tSZ component at $2000<\ell<3000$ and
\citet{sharp_09} constrain the tSZ power spectrum at $\ell\sim 4000$ to be
$\le 149\ \mu$K$^2$ at 95\% confidence level. The expected tSZ contribution
at {\it WMAP} scales is then negligible. For this reason,
we shall consider only a kSZ component. This restriction
simplifies our study since the kSZ power spectrum depends only on three parameters:
$\sigma_8$,  $\gamma T_m$ and $f_{\rm b}$. In Figure~\ref{fig:ksz_ps} we plot kSZ
power spectra for different values of $\sigma_8$ and $\gamma T_m$ with $f_{\rm b}=0.5$.
A regression fit of the variation of the maximum amplitude of the kSZ power 
spectrum with each of these three parameters permits us to write the following scaling 
relation:
\begin{equation}
A_{\rm kSZ}\sim (\gamma T_m)^{-4} \sigma_8^{14} f_{\rm b}^2~~.
\label{eq:ampksz}
\end{equation}

While the amplitude depends on the three parameters, the location of the maximum 
is only weakly dependent on $\gamma T_m$. Since $f_{\rm b}$ is a multiplicative factor 
that affects only the amplitude but not the shape or position of the maximum, 
variations on $f_{\rm b}$ and $\sigma_8$ cannot be distinguished. 
The kSZ effect is then effectively described by two parameters: one
cosmological ($\sigma_8$) and one physical ($\gamma T_m$) determining the Jeans
length. 

Figure~\ref{fig1}c also shows that most of the contribution comes
from very low redshifts. Similar scaling relations hold for
the tSZ contribution of clusters of galaxies, indicating that we
could generate arbitrary large contributions. In semi-analytical estimates,
the abundance of clusters is given by the Press--Schechter formalism
\citep{atrio_99,molnar_00}
and a large tSZ effect is obtained by arbitrarily increasing the number 
of clusters. The scaling relation given in  eq.~(\ref{eq:ampksz}) is limited 
by the validity of eq.~(\ref{eq:ksz}); if filaments overlap along the
line of sight, it overpredicts the signal. Since the kSZ effect is just
an integral along the line of sight of all electrons with a coherent
peculiar motion, an order of magnitude estimate of the effect is \citep{hogan_92}: 
\begin{equation}
(\Delta T)_{\rm kSZ}\approx 2\ \mu {\rm K} \left(\frac{\delta_{\rm b}}{50}\right) \left(\frac{f_{\rm b}}{0.5}\right)
\left(\frac{L}{30\ \rm Mpc}\right)
\left(\frac{V_{\rm B}}{600\ \rm km/s }\right) 
\label{ksz_magnitude}
\end{equation}
where $\delta_{\rm b}$ is the overdensity of a typical filament and
$L$ is the coherence scale of a motion with amplitude $V_{\rm B}$. As explained in 
\citet{atrio_08}, the distribution of the effect is rather skewed
and 9\% of all lines of sight will produce an effect 1.8--6 times
larger than the estimate from eq.~(\ref{ksz_magnitude}). In
the concordance model,  bulk flow velocities of $\sim 200$~km/s are
typical for volumes of $R\sim 100$~Mpc/h radius. Kashlinsky et al.\ (2008, 2009)
reported a bulk flow of amplitude 600--1000~km/s on a scale of $300$~Mpc/h
that could give a much larger contribution. A significant fraction of the 
temperature decrement of $-230~\mu$K detected in the intercluster medium of the 
Corona Borealis Supercluster by \citet{genova_05} could be due
to thermal and kinematic SZ contributions. However, that decrement is in 
a direction almost perpendicular to the bulk flow cited above and, because of its
orientation with respect to the line of sight, this flow
would not induce a significant kSZ contribution in Corona Borealis. Only a smaller 
kSZ contribution could exist due to peculiar motions on the scale of the supercluster itself.

\section{Markov chain Monte Carlo parameter estimation}

To explore the parameter space of $\Lambda$CDM models with and 
without a kSZ contribution, we used the April 2008 version of the 
{\sc cosmomc} package \citep{lewis_02}. This software implements an 
MCMC method that performs parameter estimation using a Bayesian approach. 
When the kSZ contribution is included, precomputed $C_\ell^{\rm kSZ}$ are added  
at each step of the chain to the theoretical power spectrum, computed with 
{\sc camb} \citep{lewis_00} for each set of $\Lambda$CDM cosmological parameters.
The model is compared with the data using the likelihood code supplied by the {\it WMAP} 
team \citep{dunkley_09}.

We considered the concordance $\Lambda$CDM model, defined by a spatially flat 
Universe with cold dark matter (CDM), baryons and a cosmological constant $\Lambda$. 
The relative contributions of these components are given
in units of the critical density: $\Omega_{\rm cdm}$, 
$\Omega_{\rm b}$ and $\Omega_\Lambda=1-\Omega_{\rm m}$, where 
$\Omega_{\rm m}=\Omega_{\rm cdm}+\Omega_{\rm b}$ is the total matter density.
The specific parameters used in the analysis were
the physical densities $\Omega_{\rm b}h^2$ and $\Omega_{\rm cdm}h^2$ where
$h=H_0/100$~km~s$^{-1}$~Mpc$^{-1}$ is the normalized Hubble constant.
In order to minimize degeneracies, instead of $h$ we used the
angular size of the first acoustic peak $\theta$, i.e. 
the ratio of the sound horizon to the angular diameter 
distance to last scattering  \citep{kosowsky_02}. We considered 
adiabatic initial conditions and
assumed an instantaneous reionization parameterized by its
optical depth to Thomson scattering up to the moment of decoupling $\tau$.  The
initial fluctuation spectrum was parameterized as a power law,
\begin{equation}
P(k) = A_{\rm S}\left(\frac{k}{k_c}\right)^{n_{\rm s}-1},
\label{eq:fluct_spec}
\end{equation}
where $A_{\rm S}$ is the amplitude at $k_c=0.05$~Mpc$^{-1}$ 
and $n_{\rm s}$ the spectral index. 
Since the effect of $f_{\rm b}$ on the kSZ power spectrum is indistinguishable from 
that of $\sigma_8$ we fix that parameter to a given value. Thus, the kSZ signal 
is defined by $\sigma_8$, which is derived from the previous cosmological parameters, 
and $\gamma T_m$. In summary, our cosmological model with the kSZ component 
has seven degrees of freedom and is described by the parameterization 
$\Theta=[\Omega_{\rm b}h^2,\Omega_{\rm cdm}h^2,100\theta,{\rm ln}(10^{10} 
A_{\rm S}),n_{\rm s},\tau,{\rm log}(\gamma T_m)]$.

In Table~\ref{tab:param_prior} we give the flat priors imposed on 
each parameter, the initial values and distribution widths.
In addition, we used a top-hat prior $10<t_0<20$~Gy for the age
of the Universe. When running the chains we included only the 
5 year {\it WMAP} data \citep{hinshaw_09} and no other
CMB or cosmological datasets.

At each step in the chain, a set of initial values for the cosmological 
parameters and for $\gamma T_m$ are generated. The $\Lambda$CDM radiation power 
spectrum and $\sigma_8$ are computed for the basic model. 
When introducing the kSZ component into the model, we would have 
to compute the values of $C_\ell^{\rm kSZ}$ at each step of 
the chain. However, as the computation is very demanding, we only 
calculate kSZ power spectra on a 2D $\sigma_8-\gamma T_m$ grid.
We built a grid with a bin width $\Delta \sigma_8=0.05$ in the interval 
$\sigma_8=[0.60,0.90]$ and $\Delta(\gamma T_m)=1\times 10^4$~K in the 
interval $\gamma T_m=[1,5]\times 10^4$~K (plus interleaved values at 
0.8, 0.9 and 1.3$\times 10^4$~K). At some random points in the parameter 
space we checked that the interpolated and exact spectrum did not differ by more than 
7\% in the interval $\ell\sim$~100--600. After a first run, we resampled our grid 
around the best-fit model. In the intervals $\sigma_8=[0.70,0.83]$ and 
$\gamma T_m=[1.05,1.95]\times 10^4$~K, the bin widths were $\Delta\sigma_8=0.01$ and 
$\Delta(\gamma T_m)=0.05\times 10^4$~K, respectively. With this resampling, 
the differences between interpolated and exact power
spectra were lower than $\sim 2$\% in the same $\ell$ range.
The kSZ power spectrum is obtained at each step of the chain by 
logarithmic interpolation of the precomputed power spectra in the four closest nodes 
of this grid. 

Initially, between two consecutive steps in the chain, we added to each 
parameter a random increment drawn from a Gaussian distribution with a 
standard deviation equal to the value listed in Table~\ref{tab:param_prior} 
multiplied by 2.4.  Since we found that $\gamma T_m$ was strongly correlated with 
$\Omega_{\rm b}h^2$ and ${\rm ln}(10^{10} A_{\rm S})$, we reran the chains using the 
eigenvalues of the covariance matrix of the parameters 
(computed with the {\sc getdist} facility which is part of the {\sc cosmomc} package) 
as distribution widths. We ran nine independent chains, with a total number of 
150\,000 independent samples, when no kSZ is included. With kSZ we fixed the baryon 
fraction in WHIM at five different values $f_{\rm b}=$0.3, 0.4, 0.5, 0.6 and 0.7 
(we also tried to constrain a model with $f_{\rm b}$ as a free
parameter, but the degeneracies did not allow us to obtain reliable results).
In each case we ran eight independent chains of similar sizes with a total 
number of samples $\gtrsim$~225\,000. We used the $R$ statistic \citep{gelman_92}
as a convergence criterion. All our parameters have $R$ well below 1.2: 
$R\approx 1.008$ for ${\rm log}(\gamma T_m)$ and $\approx$1.002 for the other parameters.

\section{Results and discussion}

In Figures~\ref{fig:1dlikes}a,b,c we plot
the mean 1D likelihoods for $\gamma T_m$, $\sigma_8$ and
the more informative rms temperature fluctuation introduced by the kSZ signal defined
as:
\begin{equation}
<\Delta {\rm T_{kSZ}}^2>^{1/2}=\left[\frac{1}{4\pi}\sum_{\ell=2}^{1000} 
(2\ell+1)C_\ell^{\rm kSZ}\right]^{1/2}~~.
\label{eq:rms_ksz}
\end{equation}
The different lines correspond to different baryon fractions. Since the kSZ 
parameters affect the amplitude of the spectrum most significantly 
(see Figure~\ref{fig:ksz_ps}) then, for each $f_{\rm b}$ the maximum 1D 
likelihood of $\gamma T_m$ shifts but the kSZ signal remains roughly constant 
(see Figure~\ref{fig:1dlikes}c).
This high degeneracy indicates that 5 year {\it WMAP} data is insensitive to the 
fraction of baryons in the WHIM and hereafter we will quote  results for 
$f_{\rm b}=0.5$.

In Figures~\ref{fig:1dlikes}d,e,f we show the mean 2D likelihoods for pairs
of parameters. Solid lines represent 1$\sigma$ and 2$\sigma$ contours, whereas
the white  
dots indicate the positions of the maximum likelihood in the full parameter space. 
Not unexpectedly, their positions are slightly shifted with respect to the 2D 
likelihood maxima. In fact, Markov chains estimate confidence intervals more precisely 
than locate the maximum of the likelihood, the bias being larger the greater the model 
dimensionality \citep{liddle_04}. Figure~\ref{fig:1dlikes}d shows a noticeable 
degeneracy between $\sigma_8$ and ${\rm log}(\gamma T_m)$. Within the 1$\sigma$ 
confidence region $A_{\rm kSZ}$ varies between $\sim 20~\mu$K$^2$ and $\sim 300~\mu$K$^2$. 
Regions with high values of $\sigma_8$ and low values of ${\rm log}(\gamma T_m)$ are 
strongly ruled out since 
they overpredict the kSZ contribution. At the 2$\sigma$ level absence of a 
kSZ contribution is compatible with the data. This is also seen in
Figures~\ref{fig:1dlikes}d,e. Models with kSZ prefer a lower value of
$\sigma_8$ (see Figure~\ref{fig:1dlikes}b), since adding kSZ requires less primordial 
CMB power. Figure~\ref{fig:1dlikes}e(f) shows that $\Omega_{\rm b}h^2$ is (inversely) proportional 
to $\langle T_{\rm kSZ}^2\rangle^{1/2}$ (log[$\gamma T_m$]). Physically, kSZ adds power mainly at 
$\ell\sim$~400--600 and the amplitude of the second acoustic peak gets reduced to fit 
the data. As a result, the best-fit value for $\Omega_{\rm b}h^2$ increases. 

The parameters that maximize the likelihood are given in Table~\ref{tab:param_results}. 
The $1\sigma$ confidence intervals have been computed from the 1D mean likelihood 
distributions. The values for the concordance model agree well with those 
given by the {\it WMAP} team \citep{dunkley_09}, indicating that we are correctly exploring 
the parameter space. The model with kSZ has larger error bars, as we are fitting the 
same data with an extra degree of freedom. The best-fit model has a kSZ power spectrum 
with a maximum amplitude of $A_{\rm kSZ}=129^{+138}_{- 52}~\mu$K$^2$ centred at 
$\ell\approx 438$, corresponding to $\gamma T_m=1.3^{+0.4}_{-0.2}\times 10^4$~K.
The rms temperature fluctuation is $\langle T_{\rm kSZ}^2\rangle^{1/2}=19^{+7}_{-9}~\mu$K, very significant 
compared with the contribution from the primordial CMB, $\langle T_{\rm CMB}^2\rangle^{1/2}= 110~\mu$K. 
None of the parameters except $\Omega_{\rm b}h^2$ differ by more than 1$\sigma$ from those 
of the concordance model. The model with kSZ gives a higher value for $\Omega_{\rm b}h^2$. 
This reinforces the slight discrepancy between the values obtained from CMB
temperature anisotropies and from BBN. As indicated above,
adding kSZ boosts the value of $\Omega_{\rm b}h^2$ since it reduces the height of
the second acoustic peak. The difference in $\sigma_8$ is also notable
even if the confidence regions overlap at the 1$\sigma$ level. 
Our estimate is closer to the values obtained from 
the number density of galaxy clusters and the optical or X-ray cluster 
mass functions. 
For example, \citet{voevodkin_04} found $\sigma_8=0.72\pm 0.04$ 
from the baryon mass fraction of a sample of 63 X-ray clusters.
After compiling cluster determinations since 2001 \citet{hetterscheidt_07} obtained
$\sigma_8=0.728\pm 0.035$, whereas from another compilation of the weak lens cosmic 
shear they found $\sigma_8=0.847\pm 0.029$.

In Figure~\ref{fig:ps}a we compare the accuracy of the fitting of the two models 
to the 5 year {\it WMAP} data. We plot the best-fit power spectra for the models 
with and without kSZ after subtracting the {\it WMAP} band powers.
The kSZ model achieves a better fit to the experimental data points
in the multipole range $\ell\sim$~300-700.
The same conclusion can be obtained from Figure~\ref{fig:ps}b, where we plot the 
ratio of the binned $\chi^2$ per $\ell$-band for the model without and 
with kSZ (note that these values were calculated, just for illustrative purposes, by 
applying the $\chi^2$ statistics to the binned best-fit theoretical power spectra, and not 
by the {\it WMAP} likelihood code). This ratio is overall $>1$ in the range 
$\ell\sim$~300-700, even though within this interval there are some particular bins where 
the concordance model produces a slightly better fit.

Introducing the kSZ component, the $\chi^2$ reduces by $\Delta\chi^2=-3.3$, from 2661.05 
to 2657.83.
Evaluating the statistical significance of this result requires taking into 
account the number of degrees of freedom, given by the difference between
the number of independent data points $N$ and of model parameters $k$. 
The {\it WMAP} likelihood is computed as a sum of different temperature and polarization 
terms \citep{dunkley_09}. In this computation 968 points correspond to the TT power 
spectrum at $\ell=33$--1000 and 427 to the TE cross-correlation at $\ell=24$--450. 
For low ($\ell\le 23$) multipoles the likelihood associated with the TT, TE, EE and 
EB correlations is evaluated directly from 1170 pixels of the temperature 
and polarization maps. In total $N=2565$. Adding the kSZ contribution
raises the number of parameters from $k=6$ to $k=7$. 
Note that the $\chi^2$ per degree of freedom also decreases when the kSZ component 
is included, from $\chi^2_{\rm dof}=1.0399$ to 1.0390. Increasing the number
of model parameters always improves the fit to a particular dataset 
but the model loses predictive power. In Bayesian statistics, 
information criteria can be used to decide whether the introduction of a new 
parameter is favored by the data. Examples are the Akaike information criterion 
\citep{akaike_74} defined as AIC~$=2k+\chi^2$ or the more conservative 
Bayesian information criterion \citep{schwarz_78} BIC~$=\chi^2+k{\rm ln}N$. 
Since we are adding a single parameter,
we obtain $\Delta$AIC~$=-1.2$, which is marginal evidence in favor of introducing 
this new parameter while $\Delta$BIC~$=4.1$ is evidence against it, reflecting the 
fact that no kSZ component is compatible with the data at the 2$\sigma$ level.
Even if at present model selection does not clearly favors the kSZ contribution,
it is a model well motivated physically and, as remarked by \citet{linder_08}, 
model selection needs to involve physical insight. A statistically more significant 
measurement would require a wider frequency coverage and angular resolution, as 
will be provided by the forthcoming {\it Planck} mission. 

\begin{figure*}
\centering
\includegraphics[width=7cm,height=6cm]{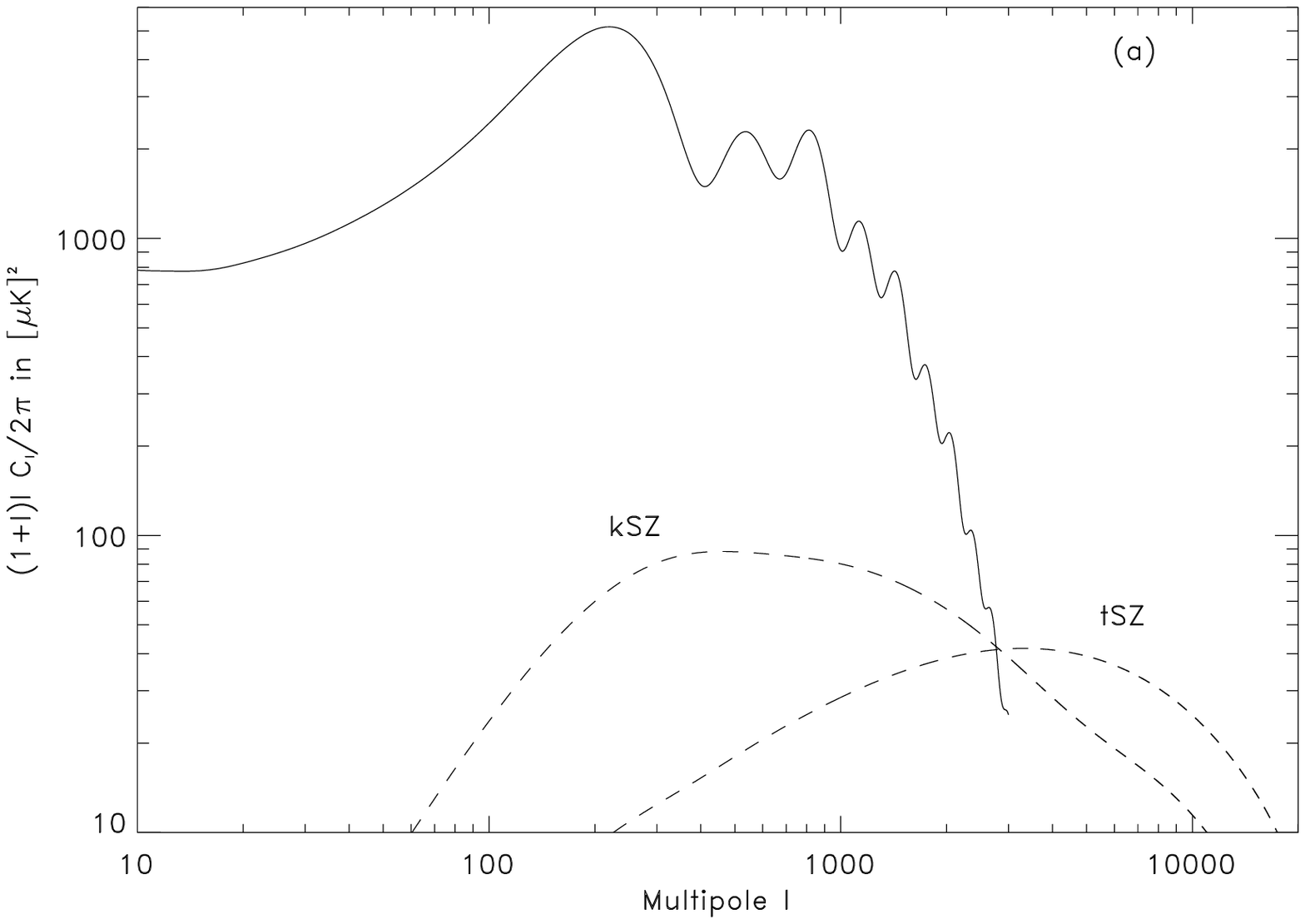}
\includegraphics[width=9cm,height=6cm]{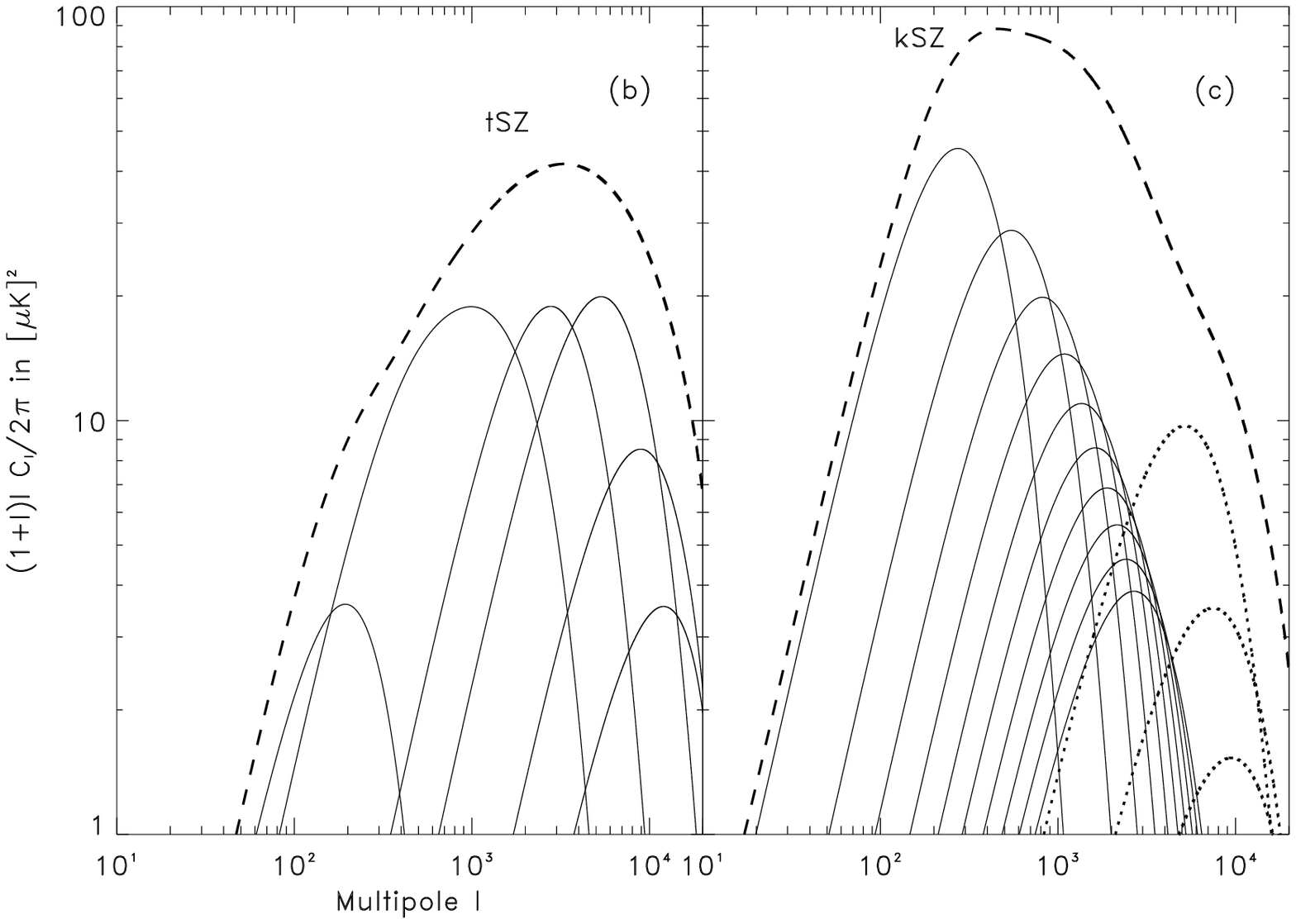}
\caption[]{(a) Intrinsic CMB radiation power spectrum (solid line) and the
IGM/WHIM tSZ and kSZ (dashed lines) contributions. (b) Contributions
of redshift intervals to the total tSZ spectrum. From left to right, 
solid lines correspond to contributions from 
the following redshift intervals: [0,0.005], [0.005,0.05], [0.05,0.1],
[0.1,0.2], [0.2,0.3] and [0.3,0.4]. (c) Redshift contribution
to the kSZ effect. Solid lines give the contribution for $z=0$ to  $0.1$ 
with spacing  $\Delta z=0.01$; dotted lines give the contributions for 
$z=0.1$ to $0.4$ with $\Delta z=0.1$. Dashed lines show the total power spectra from 
all the redshift intervals. The SZ contributions are calculated in all 
cases using $\sigma_8=0.77$, $\gamma=1.3$, $\gamma T_m=1.3\times 10^{4}$~K and $f_{\rm b}=0.5$.} 
\label{fig1}
\end{figure*}

\begin{figure*}
\centering
\includegraphics[width=8cm]{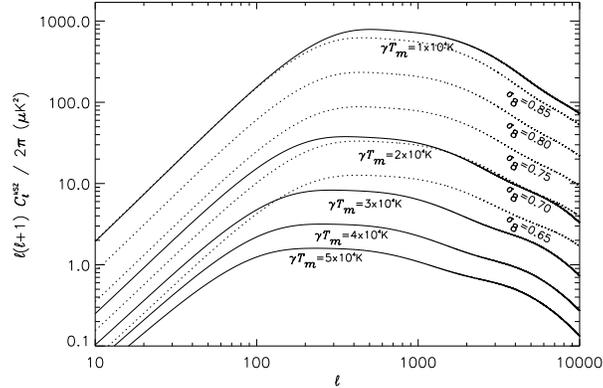}
\caption[]{Kinematic Sunyaev--Zel'dovich power spectra from WHIM for a fixed 
$\sigma_8=0.80$ and different values of $\gamma T_m$ (solid lines), and for a fixed 
$\gamma T_m=1.3\times 10^{4}$~K and different values of $\sigma_8$ (dotted lines). The fraction 
of baryons stored in the form of WHIM  has been fixed at $f_{\rm b}=0.5$ 
in all cases.
}
\label{fig:ksz_ps}
\end{figure*}

\begin{figure*}
\centering
\includegraphics[width=5.2cm]{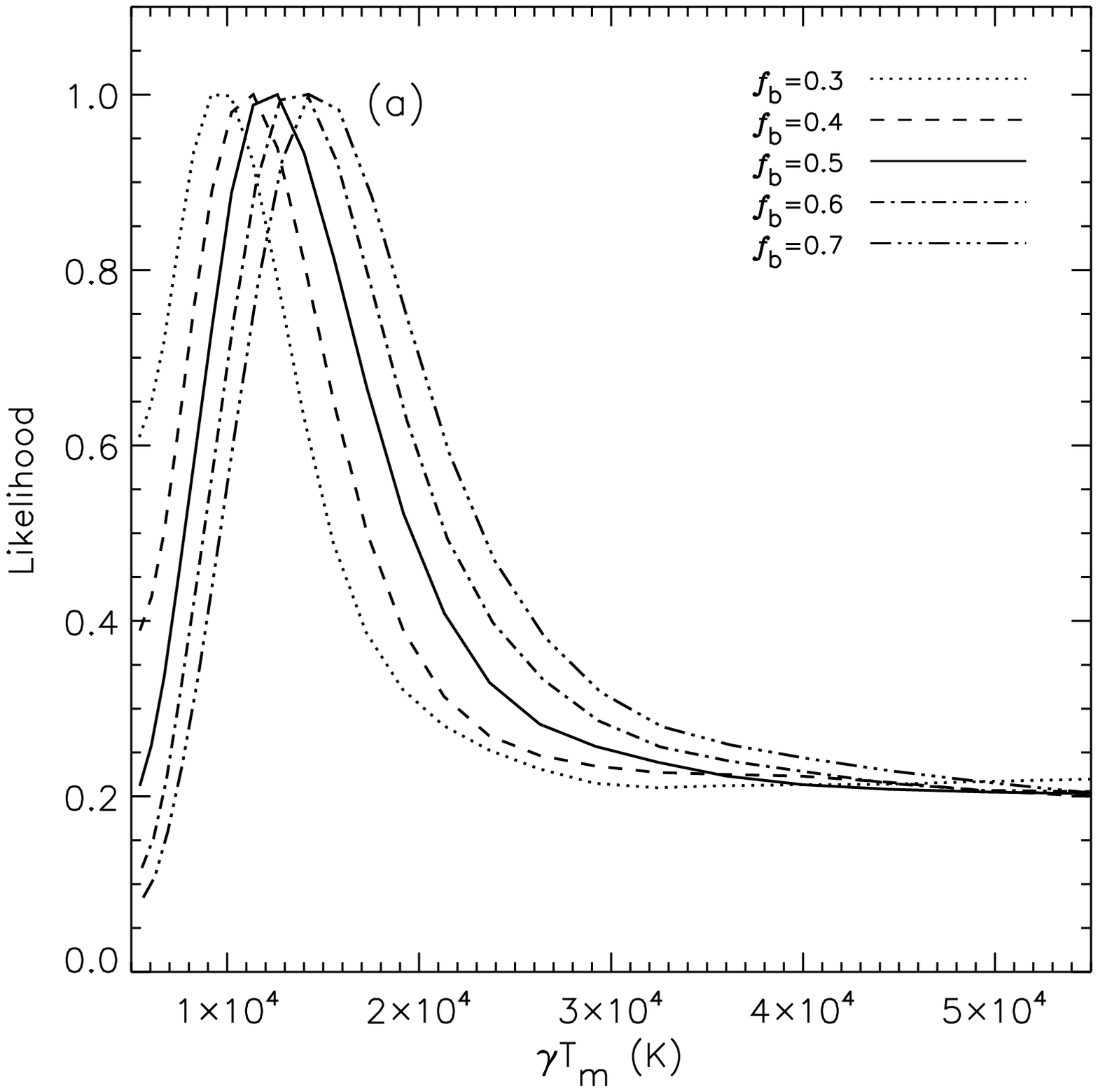}
\includegraphics[width=5.2cm]{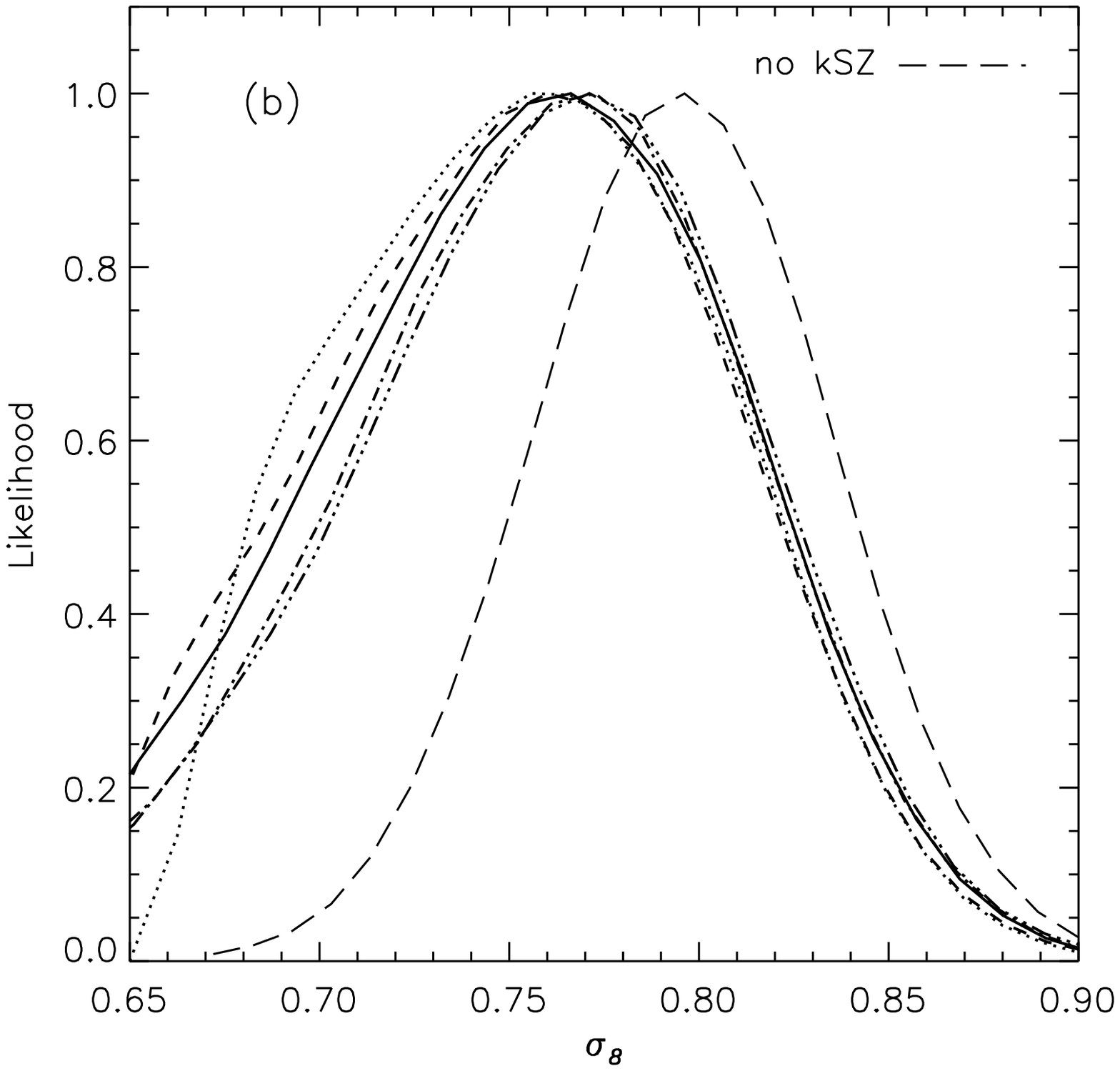}
\includegraphics[width=5.2cm]{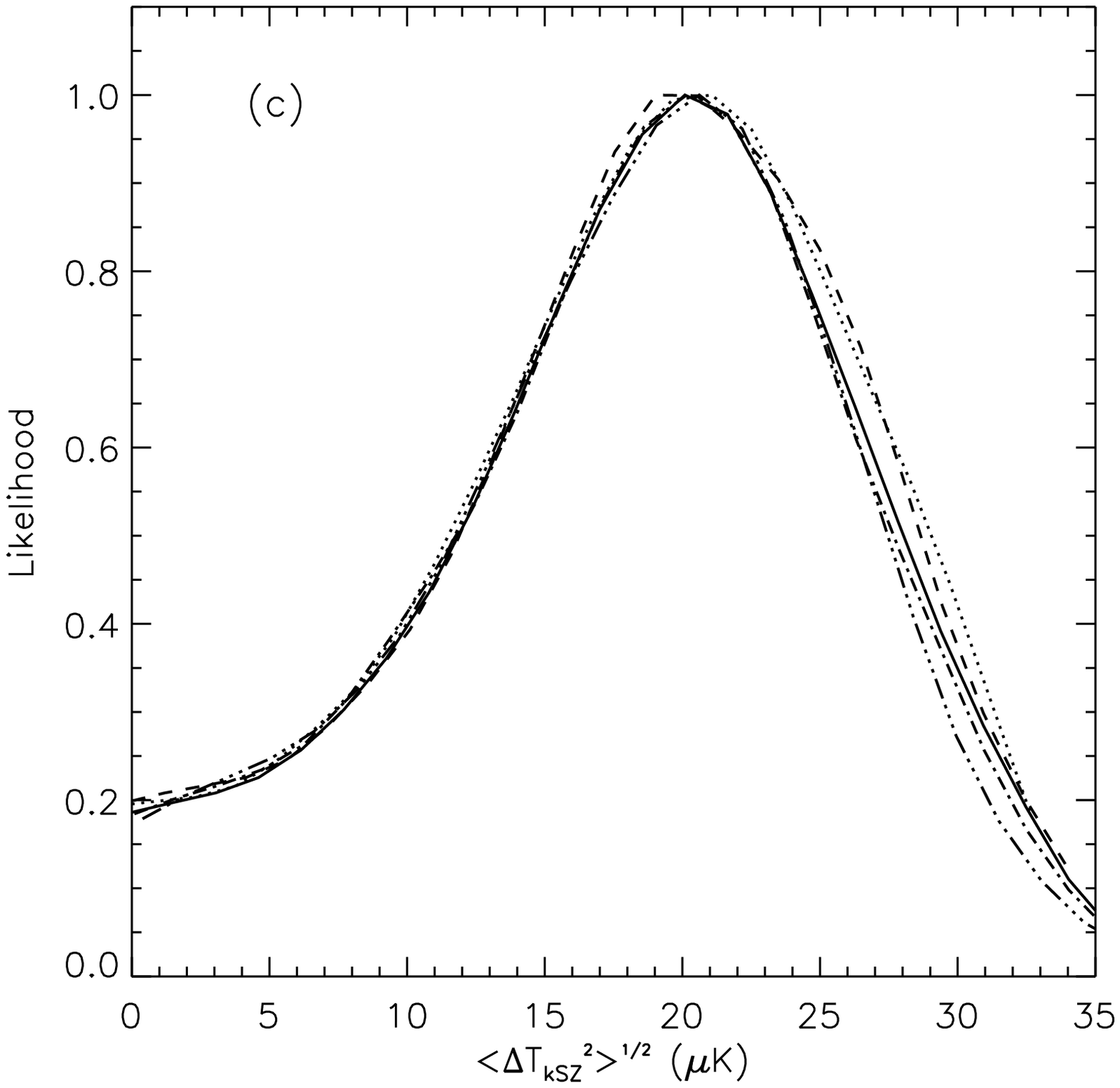}
\includegraphics[width=5.2cm]{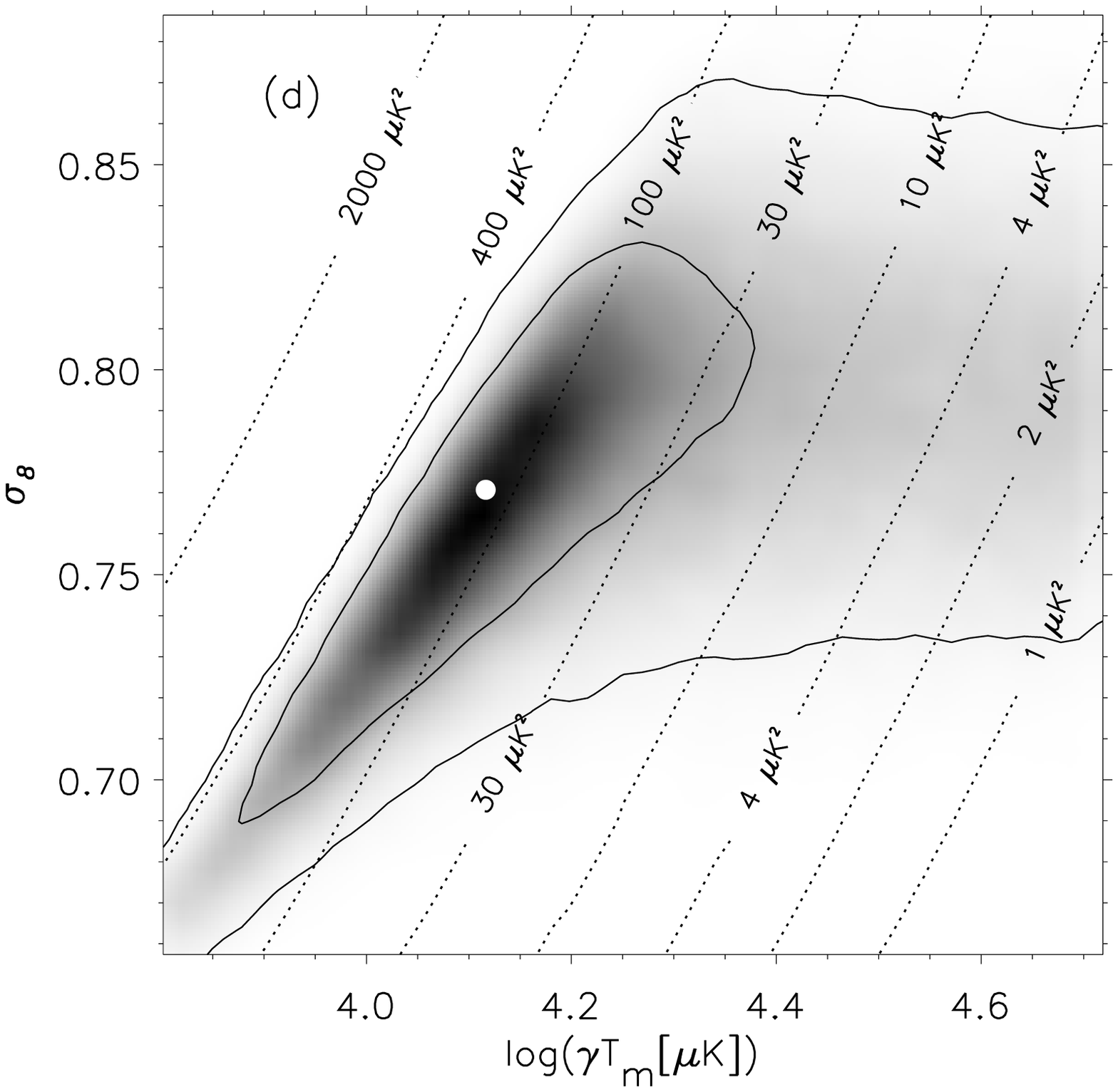}
\includegraphics[width=5.2cm]{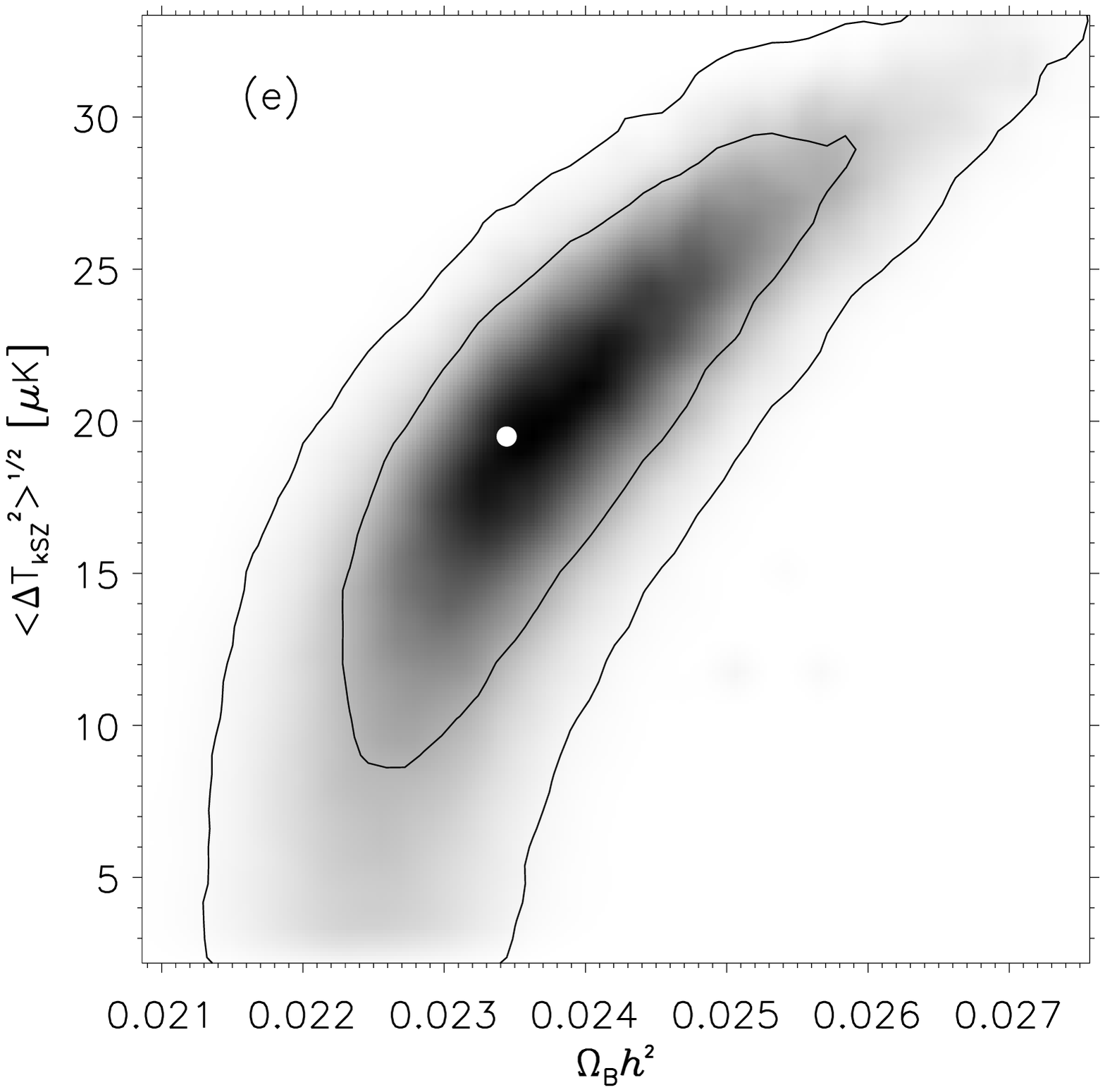}
\includegraphics[width=5.2cm]{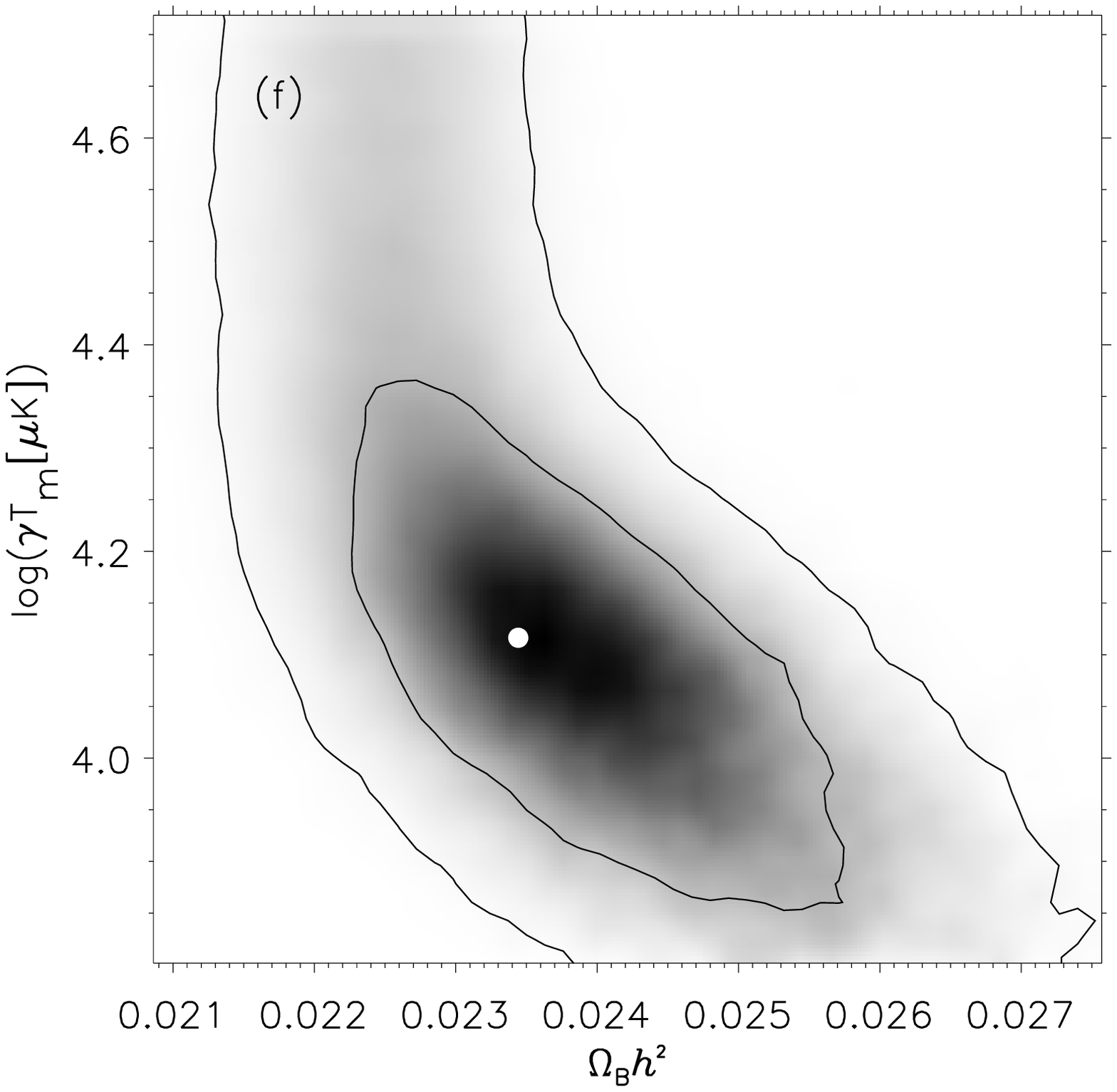}
\caption[]{Top: mean 1D likelihoods for the Jeans length $\gamma T_m$, 
$\sigma_8$ and the rms temperature anisotropy contribution of the kSZ
component. Each line correspond to a different fraction of baryons in the form of WHIM.
In (b) we also plot the likelihood for the concordance (no kSZ) model. 
Bottom: mean 2D likelihoods for three different combinations of parameters. 
Solid lines depict the 1$\sigma$ and 2$\sigma$ confidence regions, and the thick 
dots indicate the positions of the likelihood maxima found in the full parameter 
space. Dotted curves in (d) represent levels with the same kSZ amplitude.
}
\label{fig:1dlikes}
\end{figure*}

\begin{figure*}
\centering
\includegraphics[width=9cm]{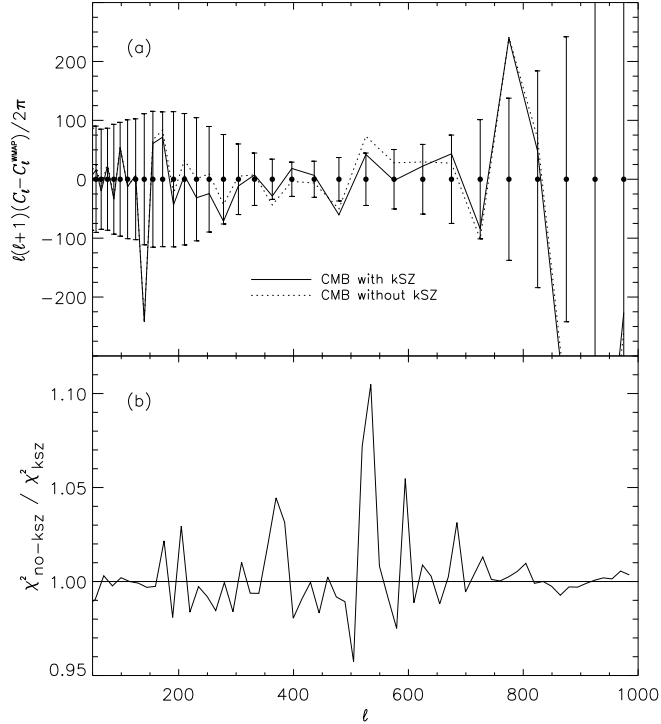}
\caption[]{Top: best-fit power spectra for the models with (solid line) and without 
(dotted line) kSZ, after subtracting the {\it WMAP} experimental band powers. 
The vertical lines represent the 1$\sigma$ error bars of the 5 year {\it WMAP} data.
Bottom: ratio of the $\chi^2$ in $\ell$-bands 
of width $\Delta\ell=15$ for a model without and with a kSZ component.
}
\label{fig:ps}
\end{figure*}

The kSZ contribution represents 3.1\% of the power measured in 5 year {\it WMAP} 
data. This contribution is larger than expected; typical filaments would give
rise to $\Delta T_{\rm kSZ}\sim$~2--5~$\mu$K (eq.~\ref{ksz_magnitude}).
We would need a more accurate (numerical) model to establish which would be
the physical conditions that give rise to the quoted kSZ signal. 
Numerical simulations over large cosmological volumes are
computationally very expensive because they require high resolution
over most of the volume. In fact, spatial resolution is usually an 
important limitation. Adaptive mesh refinement (AMR) techniques are 
efficient at describing the dynamics of gas in the high dense regions 
but their resolution falls sharply in less dense environments 
\citep{refregier_02}, while smooth particle hydrodynamics (SPH) codes 
do not yet resolve scales as small as the Jeans length \citep{bond_05}, 
that effectively dominate the WHIM SZ contribution \citep{atrio_06}.
\citet{hallman_07} used the AMR Enzo code \citep{oshea_05} to simulate 
a (512 Mpc/h)$^3$ volume including unbound bas, and found that one-third 
of the SZ flux in a 100 square-degree region comes from objects with 
masses below $5.0\times 10^{13}$~M$_\odot$ and filamentary structures made up 
of WHIM gas, a conclusion similar to that reached by \citet{hernandez_06}.
\citet{hallman_09} focused in the low-density WHIM gas by restricting their 
analysis of the same AMR simulation to regions with temperatures in the range 
$10^5-10^7$~K and overdensities $\delta <50$, even though their low mass halos 
were not yet resolved gravitationally. They computed the radiation power 
spectrum from that simulation and found a similar shape to that of 
Figure~\ref{fig1}, although with a much smaller amplitude.
The high amplitude we found could be an indirect confirmation of large scale peculiar motions
reported in \citet{kashlinsky_08} and \citet{kashlinsky_09}. If bulk flows 
of this amplitude are very common,
filaments at higher redshift could also contribute significantly, thereby increasing the 
total signal. Also, we must take into account a possible foreground contamination:
power spectra with a single maximum are also good models for Poissonianly distributed 
foreground residuals or unresolved point sources. Since the {\it WMAP} window function 
exponentially damps power at $\ell \gtrsim 600$, the convolution of a
$C_\ell$~=~const.\ spectrum with this window function peaks at $\ell<600$ and has 
a shape similar to that of the spectrum described above.  Foreground substraction can 
differ in the range 3--5~$\mu$K depending on the method \citep{ghosh_09} and
some residuals could be present on the data at that level. To
distinguish the WHIM kSZ and foreground signals we will have to include the 
frequency dependence, amplitude and location of the maximum for each component. 
In the simplest model with kSZ and one foreground, we would need four parameters to 
model both components so an analysis of all components is unfeasible with 
{\it WMAP} 3 frequency bands (Q, V and W).

To conclude,
we have explored the parameter space of the concordance model to show that the WHIM 
kSZ contribution could be as high as 3\% the total power of 5 year {\it WMAP} data. 
This large amplitude is difficult to account for in the concordance
$\Lambda$CDM model, where filaments are expected to have $\lesssim 5~\mu$K contributions. 
It could be an indication that large scale flows are rather common. 
We cannot rule out that part of this contribution could be due to 
unmasked point sources and/or foreground residuals. The {\it Planck} 
satellite, with its wider frequency coverage, lower noise and different
scanning strategy, is well suited for detecting the IGM/WHIM thermal and
kinematic contributions with much higher statistical significance and to 
distinguishing this signal from other foreground contributions. 

We are thankful to Rafael Rebolo for enlightening discussion and suggestions. 
This work is supported by the Ministerio de Educaci\'on y Ciencia and the 
``Junta de Castilla y Le\'on'' in Spain (FIS2006-05319, programa de financiaci\'on de
la actividad investigadora del grupo de Excelencia GR-234).

\begin{table*}
\begin{center}
\begin{tabular}{l c c c} 
\hline\hline
\noalign{\smallskip}
Basic & Parameter & Starting & Distribution \\
parameter & limits & points & width \\
\noalign{\smallskip}
\hline
\noalign{\smallskip}
$\Omega_{\rm b} h^2$ &                (0.005, 0.1) & 0.0223 & 0.001 \\	 
\noalign{\smallskip}
$\Omega_{\rm cdm} h^2$ & (0.01, 0.99) & 0.105  &  0.01  \\
\noalign{\smallskip}
100$\theta$&                         (0.5, 10)    & 1.04  & 0.002  \\
\noalign{\smallskip}
${\rm ln}(10^{10} A_{\rm S})$  & (2.7, 4.0)     &  3.0 & 0.01 \\
\noalign{\smallskip}
$n_{\rm s}$ &     (0.5, 1.5)   & 0.95  &  0.01 \\
\noalign{\smallskip}
$\tau$  &  (0.01, 0.8)  & 0.09  & 0.03  \\ 
\noalign{\smallskip}
${\rm log}(\gamma T_m)$  & (3.778, 4.740) &  4.096 & 0.006 \\       
\noalign{\smallskip}
\hline \hline
\end{tabular}
\end{center}
\normalsize
\medskip
\caption[tab:param_prior]{For each independent parameter, we indicate its range of variation (top-hat prior), the initital values 
and the estimated distribution widths initially used by the chain to vary each parameter.
}
\label{tab:param_prior}
\end{table*}

\begin{table*}
\begin{center}
\begin{tabular}{l c c} 
\hline\hline
\noalign{\smallskip}
Parameter & ~~~CMB alone~~~ & ~~~CMB and kSZ~~ \\
\noalign{\smallskip}
\hline
\noalign{\smallskip}
$\Omega_{\rm b} h^2$&                $0.0223^{+0.0007}_{-0.0006}$  &  $0.0234^{+0.0020}_{-0.0008}$  \\    
\noalign{\smallskip}
$\Omega_{\rm cdm} h^2$&              $0.1086^{+0.0081}_{-0.0060}$  &  $0.1096^{+0.0068}_{-0.0085}$  \\
\noalign{\smallskip}
100$\theta$&                         $1.040^{+0.004}_{-0.003}$  &  $1.042^{+0.005}_{-0.004}$  \\    
\noalign{\smallskip}
${\rm ln}(10^{10} A_{\rm S})$&       $3.06\pm 0.05$  &  $3.02^{+0.05}_{-0.08}$  \\
\noalign{\smallskip}
$n_{\rm s}$&                        $0.96\pm 0.02$  &  $0.95\pm 0.02$  \\  
\noalign{\smallskip}
$\tau$&                              $0.088^{+0.020}_{-0.019}$  &  $0.087^{+0.023}_{-0.017}$  \\ 	    
\noalign{\smallskip}
$\gamma T_{\rm m}$ (10$^4$K)&                                      &  $1.31^{+0.41}_{-0.23}$  \\
\noalign{\smallskip}
\hline
\noalign{\smallskip}
$\Omega_\Lambda$&  $0.745^{+0.027}_{-0.042}$  &  $0.752^{+0.043}_{-0.034}$  \\			       
\noalign{\smallskip}
$\Omega_{\rm m}$&   $0.255^{+0.042}_{-0.027}$  &  $0.248^{+0.034}_{-0.043}$  \\ 
\noalign{\smallskip}
$\sigma_8$&        $0.793^{+0.042}_{-0.037}$  &  $0.771^{+0.039}_{-0.069}$  \\ 	   
\noalign{\smallskip}
$z_{\rm re}$&       $10.57^{+ 1.37}_{- 1.73}$  &  $10.14^{+ 1.51}_{- 1.57}$  \\ 	  
\noalign{\smallskip}
$H_0$&         $71.63^{+ 2.85}_{- 3.12}$  &  $73.18^{+ 6.15}_{- 2.60}$ \\
\noalign{\smallskip}
$\langle T_{\rm kSZ}^2\rangle^{1/2}$ ($\mu$K)&      &  $19.5^{+ 6.8}_{- 8.7}$  \\
\noalign{\smallskip}
$A_{\rm kSZ}$ ($\mu$K$^2$)&      &  $128.7^{+138.0}_{- 52.3}$  \\ 
\noalign{\smallskip}  
\hline
\noalign{\smallskip}  
$\chi^2$    &   2661.05 & 2657.83  \\
\noalign{\smallskip}  
\hline \hline
\end{tabular}
\end{center}
\normalsize
\medskip
\caption[tab:param_results]{Best-fit parameters for the model without and with 
a kSZ contribution. These results correspond to a fraction of baryons in the form 
of WHIM of $f_{\rm b}=0.5$. The upper and lower (below the line) sets correspond 
to independent and derived parameters, respectively.
The central values have been derived from the sample with the minimum 
$\chi^2$ in the chains (also shown in the table), whereas the 1$\sigma$ 
confidence limits were derived from the 0.159 and 0.841 points of the cumulative 
probability distribution given by the 1D mean likelihoods.
}
\label{tab:param_results}
\end{table*}

\end{document}